\documentclass{sigchi-ext}

\copyrightinfo{Presented at the workshop "Should I Stay or Should I Go? Automated Vehicles in the Age of Climate Change", April 25, 2020, Honolulu, HI, USA}

\usepackage[T1]{fontenc}
\usepackage{textcomp}
\usepackage[scaled=.92]{helvet} 
\usepackage{graphicx} 
\usepackage{balance}  
\usepackage{booktabs} 
\usepackage{ccicons}  
\usepackage{ragged2e} 
\usepackage{dirtytalk} 
\usepackage[utf8]{inputenc}  
\usepackage{todonotes} 
\usepackage{float} 
\usepackage{subcaption} 
\usepackage{txfonts}

\usepackage{url}
\makeatletter
\g@addto@macro{\UrlBreaks}{\UrlOrds}
\makeatother

\usepackage{multirow} 
\usepackage{tabularx} 

\usepackage[USenglish,UKenglish]{babel}
\usepackage[nodayofweek,level]{datetime}
\usepackage[USenglish]{isodate}

\usepackage{color} 

\usepackage{enumitem} 

\usepackage[official]{eurosym}
\usepackage{siunitx} 

\usepackage[normalem]{ulem}

\usepackage{microtype}        
\usepackage[all]{hypcap}      
\usepackage{ccicons}          

\usepackage{cite} 

\def\plaintitle{Towards Reducing Energy Waste through Usage of External Communication of Autonomous Vehicles} 
  \def\plainauthor{Mark Colley, Marcel, Walch, Enrico Rukzio}

\def\plainkeywords{Autonomous vehicles; self-driving vehicles; communication; external communication; fuel efficiency.}

\title{\plaintitle}

\numberofauthors{3}
\author{
  \alignauthor{%
    \textbf{Mark Colley}\\
    \affaddr{Institute of Media Informatics} \\
    \affaddr{Ulm University, Germany} \\
    \email{mark.colley@uni-ulm.de} } \vfil 
    \alignauthor{%
    \textbf{Marcel Walch}\\    
    \affaddr{Institute of Media Informatics} \\
    \affaddr{Ulm University, Germany} \\
    \email{marcel.walch@uni-ulm.de} }\vfil 
    \alignauthor{%
    \textbf{Enrico Rukzio}\\    
    \affaddr{Institute of Media Informatics} \\
    \affaddr{Ulm University, Germany} \\
    \email{enrico.rukzio@uni-ulm.de} }
    }

\definecolor{linkColor}{RGB}{6,125,233}
\hypersetup{%
  pdftitle={\plaintitle},
  pdfauthor={\plainauthor},
  pdfkeywords={\plainkeywords},
  bookmarksnumbered,
  pdfstartview={FitH},
  colorlinks,
  citecolor=black,
  filecolor=black,
  linkcolor=black,
  urlcolor=linkColor,
  breaklinks=true,
}

\begin{document}


\maketitle

\RaggedRight{} 

\begin{abstract}
Automated vehicles can implement strategies to drive with optimized fuel efficiency. Therefore, automated driving is seen as a major advancement in tackling climate change. 
However, with automated vehicles driving in cities and other areas rife with other road users such as human drivers, pedestrians, or cyclists, there is the potential for \say{stop-and-go} traffic. This would greatly diminish the possibility of automated vehicles to drive fuel-efficient.
We suggest using \textit{external communication of automated vehicles} to aid in ecological driving by providing clues to other road users to show the intent and therefore ultimately enable smoother traffic.

\end{abstract}



\newpage
\section{Introduction}\label{sec:introduction}
Road traffic is a major contributor to the global emission of carbon dioxide. In Europe, for example, road transportation accounted for approximately 21\% of the carbon dioxide emissions~\cite{euco2}. Traffic congestion leads to increased usage of fuel~\cite{walraven2016traffic}. Therefore, ways to reduce congestion and, in general, to improve traffic flow, have to be found.
Autonomous traffic is expected to change traffic fundamentally~\cite{fagnant2015preparing, aria2016investigation}. 
Platooning is expected to improve traffic flow and fuel efficiency by reducing air resistance~\cite{fernandes2012platooning}. For this, real-time capable technologies are needed. First tests have shown the benefit of platooning while being not as efficient as hoped~\cite{heise2019platooning}. Smoothing of traffic flow in general was assessed as resulting in reduced emissions~\cite{stathopoulos2003induced}.

In recent years, the external communication of autonomous vehicles (AVs) has become a novel field of research. This research is targeted towards replacing interpersonal communication used today between human drivers and vulnerable road users (VRUs) towards improving traffic safety for people with disabilities~\cite{colley2020towards}. Various areas within this field have been explored: external communication modalities such as displays~\cite{RefWorks:doc:5cf7ad8de4b06bba938e0112}, LED strips~\cite{RefWorks:doc:5cf7ad8de4b06bba938e0112, RefWorks:doc:5cf8aa47e4b006bc06a90b0a}, movement patterns~\cite{RefWorks:doc:5ce536a5e4b076262b5839fa}, projections~\cite{ackermann2019experimental} auditory or tactile cues~\cite{mahadevan2018communicating}, as well as combinations thereof~\cite{mahadevan2018communicating} and enhancement of the infrastructure~\cite{RefWorks:doc:5cd92775e4b0487541989799}, trust challenges~\cite{hollander2019overtrust}, and (some) legal issues~\cite{inners2017beyond}.

While first tests with the mentioned technologies are promising, the hoped-for advances were not met. With the advancement of AVs, there is the fear of having even more miles travelled~\cite{fagnant2015preparing} on the street as these vehicles are able to look for parking spots in ever greater distance to the point of destination.

We introduce external communication of AVs towards other road users as one idea to smooth traffic flow. External communication of the intent~\cite{iso23049} can lead towards better understanding of the actions of AVs for pedestrians and therefore for fewer interruptions.
In this position paper, we outline classic traffic characteristics and how we believe external communication could influence these.

\section{External Communication of Autonomous Vehicles}
AVs do not need a human passenger to be present. Therefore, situations in which pedestrians and human drivers today rely on gestures or eye-contact could become very challenging. Recent research projects try to overcome these challenges via substitution of interpersonal communication with communication between AV and VRU (e.g.~\cite{locken2019should, colley2020towards, mahadevan2019av, deb2019comparison, dey2018interface}).
Müller et al.~\cite{muller2016social} already showed that AVs could be viewed as social actors in traffic. We report on related work in external communication of AVs to VRUs and other human drivers.

\subsection{External Communication Towards VRUs}
Various works have investigated communication between AVs and pedestrians. While there is still debate about the necessity of external communication~\cite{moore2019case}, many studies showed a benefit of external communication~\cite{colley2020towards, mahadevan2019av}. The current ISO standard~\cite{iso23049} advices to communicate intent towards VRUs instead of giving advice (\say{Safe to cross}) or even commands (\say{Cross}). However,other studies found that giving an unambiguous command was rated best~\cite{hudson2018pedestrian}.
The presented communication concepts have been categorized in various ways: by their complexity~\cite{locken2019should}, by their modality~\cite{mahadevan2018communicating, colley2020towards}, their information content~\cite{colley2020towards}, the locus of the communication~\cite{mahadevan2018communicating}, and their inclusiveness of people with visual impairments~\cite{colley2019including}. Different addressees have been compared, for example adults and children~\cite{deb2019comparison}.
Numerous technologies were proposed as described in the~\nameref{sec:introduction}.
However, this research only focused on a pedestrian who wants to cross a street.
\break
There is little literature about communication of AVs towards cyclists, however, there is a patent from Lyft~\cite{matthiesen2018autonomous} showing the AV communicating to the cyclist that it is \say{safe to cross}.

\subsection{External Communication towards Human Drivers}
Communicating the intent and explaining why an AV behaved the way it did was already investigated~\cite{koo2015did}. However, to our best knowledge, there is no current research on how an AV could communicate its intent to other drivers other than concepts from industry such as the F 015~\cite{f015} which, for example, indicates towards other human drivers to stop.
A video showcase presented by Wang et al.~\cite{wang2017hud} used a Head-Up Display (HUD) to show information about the other drivers and their intentions, e.g., driving to an airport. However, this was not in an autonomous context and other vehicles had to be equipped with visual markers. 
The aforementioned patent of Lyft~\cite{matthiesen2018autonomous} also showed that by using the side of the vehicle as a display, it can communicate that it is \say{yielding} at an intersection.

\section{Fuel Efficiency via External Communication}
Improving driver efficiency could lead (without AVs) to a reduction of about 20\%~\cite{gonder2012analyzing}. This would be achieved by avoiding stop-and-go, unnecessary idling and optimized acceleration and speed~\cite{gonder2012analyzing}.\break
With AVs, the same negative behavior has to be avoided. Optimized acceleration and speed will most likely be realized via algorithms, but stop-and-go and idling could be affected by humans.
Fuel efficiency could be decreased because of frequent slowing down and stops of AVs. In cities, a major reason for this could be people crossing in front of the vehicle. Millard-Ball already discussed this anticipated phenomenon as \say{crosswalk chicken}~\cite{millard2018pedestrians}. He claims that people will frequently cross streets in front of an AV as the vehicles will most likely be programmed to drive extra cautiously~\cite{millard2018pedestrians}.
This could be avoided through displays either by communicating early that a pedestrian can cross (cf.~\cite{locken2019should}) to avoid the need for a full stop or to to prevent pedestrians to cross too often (i.e., by displaying a warning). Millard-Ball~\cite{millard2018pedestrians} already states that an AV has to have some uncertainty whether it will stop to avoid too frequent crossings of pedestrians. Besides crossings, this could also be useful in crowded spaces such as parking lots or pedestrian areas.\break
With Avs, idling will become less. However, situations may arise in which a vehicle is waiting for a passenger (e.g., talking to other person). Depending on the settings, this could lead to unused energy in the form of heating. The vehicle could remind the passenger of this in a suitable way.

With regard to communication towards other human drivers, the intention of the AV can be communicated for example via external displays or via Vehicle-To-X (V2X) technology~\cite{liu2013opportunities}. A message showing that a (1) following vehicle has to slow down or that (2) a vehicle is about to turn (clearer than flashing lights) or (3) that a vehicle is yielding~\cite{matthiesen2018autonomous} can aid the other (manual) driver to avoid fuel inefficient maneuvers. To increase awareness towards fuel saving, an AV could also show on its display that it is currently driving in an eco-friendly way, therefore trying to persuade other to do the same, for example on a highway.

\section{Conclusion}
AVs are expected to reduce energy waste by driving more efficient~\cite{gonder2012analyzing}. However, there are unforeseeable events that could diminish this effect, such as pedestrians crossing frequently in front of these vehicles, causing a lot of stops~\cite{millard2018pedestrians}. We suggest to use external communication of AVs to reduce this. Additionally, we hope to encourage other human drivers to drive more efficiently by providing information about an eco-friendly driving style.
Moreover, we hope to spark a discussion about novel ways to reduce energy waste besides typical (technical or regulatory) approaches. Furthermore, we hope to gain knowledge from other domains and potentially start fruitful collaborations.


\section{Acknowledgements}
This work was conducted within the project 'Interaction between automated vehicles and vulnerable road users' (Intuitiver) funded by the Ministry of Science, Research and the Arts of the State of Baden-Württemberg.

\balance{} 

\bibliographystyle{SIGCHI-Reference-Format}
\bibliography{sample.bib}

\end{document}